\documentclass[runningheads]{llncs}
\usepackage{graphicx}
\usepackage{graphicx}
\usepackage{algorithm}
\usepackage{algorithmic}
\usepackage{amsmath,amssymb}
\usepackage{mathrsfs}
\usepackage[section]{placeins}
\usepackage{subfigure}
\usepackage{booktabs}
\usepackage{multirow}
\usepackage{float}
\usepackage[dvipsnames]{xcolor}
\usepackage{wrapfig}
\usepackage{bbding}
\usepackage{newclude}

\begin{document}
%
\title{MambaMIR: An Arbitrary-Masked Mamba for Joint Medical Image Reconstruction and Uncertainty Estimation}

\titlerunning{MambaMIR}

\author{
Jiahao Huang\inst{1, 2, 3} \Envelope \and
Liutao Yang\inst{1, 2, 4} \and
Fanwen Wang\inst{1, 2, 3} \and
Yang Nan\inst{1, 2} \and
Angelica I. Aviles-Rivero\inst{5} \and
Carola-Bibiane Sch{\"o}nlieb\inst{5} \and \\
Daoqiang Zhang\inst{4} \and
Guang Yang\inst{1, 2, 3} \Envelope
}

\authorrunning{J. Huang et al.}

\institute{
Bioengineering Department and Imperial-X, Imperial College London, London, UK \\
\email{\{j.huang21, g.yang\}@imperial.ac.uk} 
\and National Heart and Lung Institute, Imperial College London, London, UK 
\and Cardiovascular Research Centre, Royal Brompton Hospital, London, UK 
\and College of Computer Science and Technology, Nanjing University of Aeronautics and Astronautics, China
\and Department of Applied Mathematics and Theoretical Physics, University of Cambridge, UK}


\maketitle              
\begin{abstract}
The recent Mamba model has shown remarkable adaptability for visual representation learning, including in medical imaging tasks. 
This study introduces MambaMIR, a Mamba-based model for medical image reconstruction, as well as its Generative Adversarial Network-based variant, MambaMIR-GAN.
Our proposed MambaMIR inherits several advantages, such as linear complexity, global receptive fields, and dynamic weights, from the original Mamba model. 
The innovated arbitrary-mask mechanism effectively adapt Mamba to our image reconstruction task, providing randomness for subsequent Monte Carlo-based uncertainty estimation.
Experiments conducted on various medical image reconstruction tasks, including fast MRI and SVCT, which cover anatomical regions such as the knee, chest, and abdomen, have demonstrated that MambaMIR and MambaMIR-GAN achieve comparable or superior reconstruction results relative to state-of-the-art methods. 
Additionally, the estimated uncertainty maps offer further insights into the reliability of the reconstruction quality. 
The code is publicly available at https://github.com/ayanglab/MambaMIR. 

\end{abstract}

\section{Introduction}

\begin{figure}[t]
    \centering
    \includegraphics[width=\textwidth]{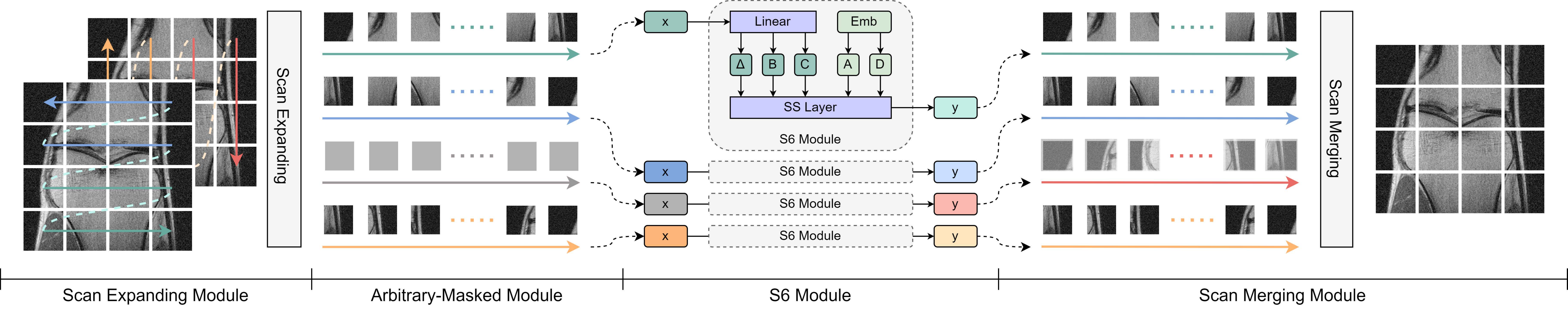}
    \caption{
    The detailed structure of the proposed Arbitrary-Masked S6 (AMS6) Block.
    }
    \label{fig:FIG_AMS6-BLOCK-DETAIL}
\end{figure}

Medical imaging plays an essential role in supporting the diagnostic and therapeutic decision-making process in contemporary clinical settings. 
Image reconstruction stands as one of the most fundamental and pivotal components of medical imaging~\cite{Wang2020Deep}. 
High-quality and high-fidelity reconstructed medical images ensure the accuracy and effectiveness of subsequent disease detection, diagnosis, and treatment planning, thereby reducing potential risks to patient health.

Magnetic Resonance Imaging (MRI) and X-ray Computed Tomography (CT) are two representative medical imaging tools. 
Magnetic resonance imaging serves as a crucial non-invasive diagnostic tool, offering high-resolution and reproducible assessments of both structural and functional information, without exposure to radiation. 
Fast MRI is widely utilised to produce MR images from sub-Nyquist sampled \textit{k}-space measurements, aiming to speed up the inherently slow data acquisition process~\cite{Liang2020Deep, Hammernik2022Physics, Huang2024Data}. This accelerated acquisition process can effectively reduce artifacts from voluntary and involuntary physiological movements during scanning. 
X-ray Computed Tomography is renowned for its ability to generate high-quality and detailed images. However, it also raises significant concerns regarding potential cancer risks due to radiation exposure~\cite{shah2008alara}. 
Sparse-view CT (SVCT), which reconstructs images from far fewer projection views, can effectively reduce the radiation dose. Moreover, in applications such as C-arm CT or dental CT, where scanning time is predominantly constrained by the slower flat-panel detector rather than the gantry's mechanical speed, sparse-view CT emerges as a promising solution to expedite the scanning process~\cite{pan2009commercial}. However, reducing the number of projection views during scanning can introduce severe artefacts in reconstructed images, which can significantly undermine clinical diagnosis. Therefore, for fast MRI, SVCT, and other medical image reconstruction tasks, a key research topic and challenge is to develop effective, efficient, and reliable reconstruction models.
Therefore, for fast MRI, SVCT as well as other medical image reconstruction task, a key research topic and challenge is to develop effective, efficient and reliable reconstruction model.

\begin{figure}[t]
    \centering
    \includegraphics[width=\textwidth]{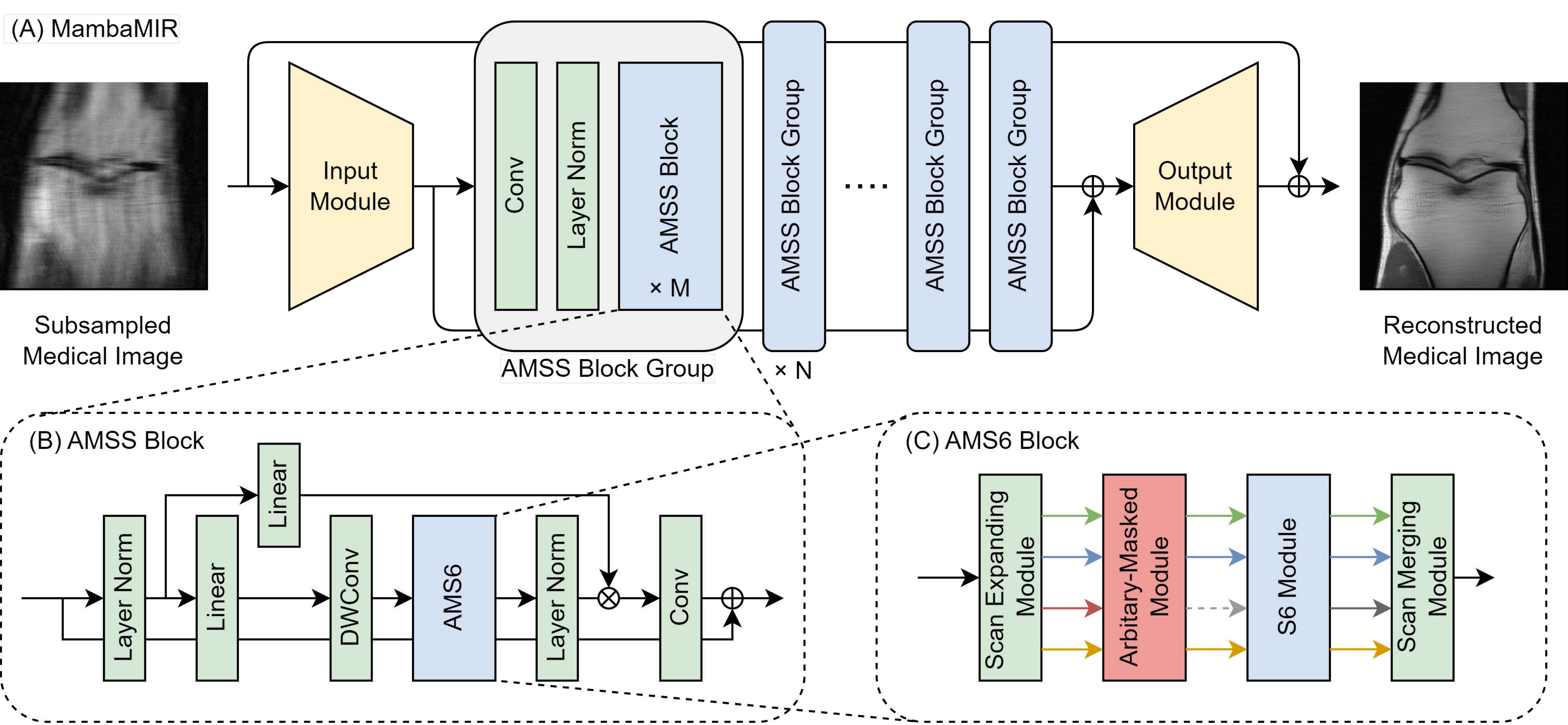}
    \caption{
    (A) The architecture of the proposed MambaMIR. The mamba is composed of are $m$ Arbitrary-Masked State Space (AMSS) Block Groups, where each group contains $n$ AMSS Block;
    (B) The structure of the AMSS Block;
    (C) The structure of the Arbitrary-Masked S6 (AMS6) Block.
    }
    \label{fig:FIG_ARCHITECTURE}
\end{figure}

The rapid advancement of artificial intelligence has propelled the development and widespread application of deep learning (DL)-based medical image reconstruction. Convolutional Neural Networks (CNNs) and Vision Transformers (ViTs)~\cite{Dosovitskiy2020ViT} represent two mainstream models that have achieved remarkable success in various computer vision tasks and are widely utilised in the medical imaging field.
CNNs, with their hierarchical architecture and inductive biases, excel at extracting features, especially in identifying local patterns. The shared weight mechanism renders them more parameter-efficient than multilayer perceptrons (MLPs). Various CNN-based medical image reconstruction models have been proposed for different modalities, including MRI~\cite{Schlemper2017D5C5, Aggarwal2019MoDL}, CT~\cite{jin2017deep, Gupta2018CNN}, Ultrasound~\cite{Zhou2018High}, and Positron Emission Tomography (PET)~\cite{Gong2019PET}. However, CNNs typically exhibit local sensitivity and a lack of long-range dependency, which limits their ability to contextualise global features.
Vision Transformers~\cite{Dosovitskiy2020ViT}, characterised by their large receptive fields and global sensitivity, often outperform CNNs in capturing extensive contextual information. Nonetheless, their significant computational demand, driven by the self-attention mechanism's quadratic complexity with respect to sequence length, limits their practicality for medical image reconstruction, where images often have high resolution. Moreover, utilising larger patch sizes may compromise the reconstruction of detailed information~\cite{Liang2021SwinIR}. Recent Transformer-based models for medical image reconstruction have sought to mitigate these limitations by:
1) Adopting a trade-off strategy that applies the self-attention mechanism within shifting windows rather than across the entire feature map~\cite{Huang2022SwinMR};
2) Constructing hybrid models that incorporate CNNs~\cite{Chen2021TransUNet} or Swin Transformers~\cite{Liu2021Swin}, applying ViT blocks only within deep, low-resolution latent spaces~\cite{Chen2021TransUNet, Huang2022SDAUT}.

Mamba~\cite{Gu2023Mamba} is an emerging sequence model derived from natural image processing tasks that has been proposed for learning visual representations~\cite{Zhu2024VisionMamba, Liu2024VMamba}. It has been further applied to downstream tasks in medical imaging, such as segmentation~\cite{Ma2024UMamba, Ruan2024VMUNet} and detection~\cite{Gong2024nnMamba}.

Motivated by the success of Mamba for Vision~\cite{Zhu2024VisionMamba, Liu2024VMamba}, we propose MambaMIR, a Mamba-based model for joint medical image reconstruction and uncertainty estimation. Furthermore, we explore its variants based on the Generative Adversarial Network (GAN), namely MambaMIR-GAN. 
MambaMIR inherits several advantages from the original Mamba model, including linear complexity, global receptive fields, and dynamic weights. These features are particularly well-suited for medical image reconstruction tasks, which often require processing long sequences (large spatial resolutions) to preserve detailed information and ensure global sensitivity through long-range dependency.
We introduce a novel arbitrary-mask mechanism (Fig.~\ref{fig:FIG_AMS6-BLOCK-DETAIL}) to adapt Mamba to our image reconstruction task, providing randomness for subsequent Monte Carlo-based uncertainty estimation by randomly masking out redundant image scan sequences. Experiments conducted on various medical image reconstruction tasks, including fast MRI and SVCT, covering anatomical regions such as the knee, chest, and abdomen, have demonstrated that MambaMIR achieves comparable or superior reconstruction results relative to state-of-the-art methods. Additionally, the estimated uncertainty maps offer further insights into the reliability of the reconstruction quality.

Our main contributions are summarised as follows:
\begin{itemize}
    \item[$\bullet$] 
    Proposal of an innovative Mamba-based model for joint medical image reconstruction and uncertainty estimation. To the best of our knowledge, MambaMIR is the first Mamba-based model applied to medical image reconstruction. We also explore its GAN variants, i.e., MambaMIR-GAN.
    \item[$\bullet$]
    Introduction of an Arbitrary-Mask mechanism to adapt non-causal medical image data for our model, providing randomness for uncertainty estimation by arbitrarily masking out redundant scan sequences.
    \item[$\bullet$]
    Demonstration of promising results by MambaMIR across different medical image modalities and anatomical regions.
\end{itemize}

\section{Methodology}

\subsection{Preliminaries}

\subsubsection{Medical Image Reconstruction.}
The forward acquisition process for medical images is described by:
\begin{equation}\label{eq:forward}
\begin{aligned}
\mathbf{y}=\mathbf{A} \mathbf{x} + \mathbf{n},
\end{aligned}
\end{equation}
where $\mathbf{x} \in \mathbb{C}^n$ represents the image of interest, $\mathbf{y} \in \mathbb{C}^m$ denotes the corresponding measurements, and $\mathbf{n} \in \mathbb{C}^m$ is the inevitable noise encountered during the measurement process. 
Depending on the type of medical imaging, the forward operator $\mathbf{A}$ can vary. For fast MRI, $\mathbf{A}$ can be a subsampled discrete Fourier transform $\mathcal{F}_{\Omega}: \mathbb{C}^n \rightarrow \mathbb{C}^m$, sampling the \textit{k}-space locations as specified by $\Omega$. For sparse-view CT, $\mathbf{A}$ is represented by the Radon transform $\mathcal{R}_{\Gamma}: \mathbb{C}^n \rightarrow \mathbb{C}^m$, projecting targets into a sinogram under a selected set of imaging angles $\Gamma$.
Generally, the goal of the reconstruction stage is to recover the ground truth $\mathbf{x}$ from the undersampled measurements $\mathbf{y}$. This process can be formulated as an inverse problem:
\begin{equation}\label{eq:inverse}
\begin{aligned}
\hat{\mathbf{x}}=\arg \min _{\mathbf{x}} \frac{1}{2}\|\mathbf{A} \mathbf{x} - \mathbf{y}\|_2^2+\lambda \mathcal{R}(\mathbf{x}),
\end{aligned}
\end{equation} 
where $\mathcal{R}$ represents a class of regularisers, and $\lambda$ is a balancing parameter. This formulation aims to minimise the discrepancy between the measured and predicted data while incorporating regularisation to impose prior knowledge or desired properties on the solution.

\subsubsection{State Space Model and its Discretisation.}

State Space Models (SSMs) have emerged as a foundational framework for the analysis of sequence data, inspired by systems theory which describes a system's dynamics through its state transitions~\cite{Gu2021Efficiently}. 
SSMs are typically characterised as linear, time-invariant systems that map an input sequence $x(t) \in \mathbb{R}^L$ to an output sequence $y(t) \in \mathbb{R}^L$ through a series of hidden states $h(t) \in \mathbb{R}^N$. These models can be expressed using linear ordinary differential equations:
\begin{equation}\label{eq:SSM_C}
\begin{gathered}
h^{\prime}(t) = \mathbf{A} h(t) + \mathbf{B} x(t), \quad y(t) = \mathbf{C} h(t) + \mathbf{D} x(t),
\end{gathered}
\end{equation}
where $\mathbf{A} \in \mathbb{R}^{N \times N}$, $\mathbf{B} \in \mathbb{R}^{N \times 1}$, and $\mathbf{C} \in \mathbb{R}^{1 \times N}$ represent the learnable parameters, with $\mathbf{D} \in \mathbb{R}^{1}$ typically denoting a residual connection.

The structured state space sequence models (S4)~\cite{Gu2021Efficiently} and more recent Mamba~\cite{Gu2023Mamba} are based on a discretised version of these continuous models:
\begin{equation}\label{eq:SSM_D}
\begin{gathered}
h_k =\bar{\mathbf{A}} h_{k-1}+\bar{\mathbf{B}} x_k, \quad y_k =\bar{\mathbf{C}} h_k+\bar{\mathbf{D}} x_k, \\
\bar{\mathbf{A}} = e^{\mathbf{\Delta} \mathbf{A}}, \quad
\bar{\mathbf{B}} = (e^{\mathbf{\Delta} \mathbf{A}}-I) \mathbf{A}^{-1} \mathbf{B} \approx \mathbf{\Delta} \mathbf{B}, \quad
\bar{\mathbf{C}} =\mathbf{C}, \quad
\bar{\mathbf{D}} =\mathbf{D},
\end{gathered}
\end{equation} 
where $\bar{\mathbf{A}}$, $\bar{\mathbf{B}}$, $\bar{\mathbf{C}}$, $\bar{\mathbf{D}}$ are the discretised parameter, transformed by a timescale parameter $\mathbf{\Delta}$.

\subsubsection{Vision Mamba.}

Mamba introduces a novel approach in the landscape of State Space Models (SSMs) with its Selective Structured State Space Sequence Models incorporating a Scan (S6)~\cite{Gu2023Mamba}. This innovation allows for the dynamic parameterisation of the SSM, with parameters $\bar{\mathbf{B}}$, $\bar{\mathbf{C}}$, and $\mathbf{\Delta}$ being derived directly from the input data $x$, enabling an input-specific adaptation of the model.

Characterised by linear complexity and enhanced through hardware-aware optimisations, Mamba exhibits superior efficiency in managing long sequence modelling. This efficiency positions Mamba as a viable contender to the prevalent self-attention mechanisms found in Transformers, particularly for tasks involving the processing of high-resolution visual data.

Despite its advantages, S6's sequential input processing presents a limitation in the context of information integration, as it can only incorporate data that has been sequentially processed. This sequential nature aligns well with Natural Language Processing (NLP) tasks, which are inherently temporal. However, challenges arise when applying S6 to computer vision tasks, where data types are not strictly sequential.

To address these challenges, innovative solutions have been proposed. These include conducting multiple scans across different spatial directions~\cite{Zhu2024VisionMamba, Liu2024VMamba} and adopting position embedding techniques similar to those used in Transformers~\cite{Zhu2024VisionMamba}. Such adaptations aim to overcome the directional sensitivity issue, enhancing the model's applicability to a broader range of tasks, including those in the domain of computer vision.


\subsection{Arbitrary-Masked S6 Block with Monte Carlo Dropout}

Building upon the work of Liu et al.~\cite{Liu2024VMamba}, we introduce the Arbitrary-Masked S6 Block (AMS6 Block), a novel component designed to enhance the capability of State Space Models (SSMs) for visual data processing. The AMS6 Block integrates four key modules: the Scan Expanding Module, the Arbitrary-Masked Module, the S6 Module, and the Scan Merging Module, as depicted in Fig.~\ref{fig:FIG_AMS6-BLOCK-DETAIL}. This block serves as the foundational element of the Arbitrary-Masked State Space Block (AMSS Block).

The Scan Expanding Module unfolds image patches along either rows or columns, starting from either the left-top or the right-bottom patch. This process converts a single image into four ordered sequences, effectively expanding the image data by a factor of $4\times$. 
By this way, one image is $4\times$ expanded and redundant, since every scans carry the same information and only differ in the scanning direction. 

In the Arbitrary-Masked Module, we introduce randomness by arbitrarily masking scans, leveraging the scanning redundancy. This is achieved by setting to zero the values in $s$ of four randomly chosen scans ($s \in \{0, 1, 2, 3\}$), thereby maintaining the original matrix shape while selectively obscuring information. 
This method draws inspiration from Monte Carlo dropout techniques~\cite{Gal2016Dropout}, utilising randomness to generate a distribution of predictions (or reconstructions) from a single subsampled measurement during inference. The variance or standard deviation within this distribution provides a measure of the model's uncertainty, incorporating randomness in both training and inference phases.

The core of the AMS6 Block is the S6 module, which processes the aforementioned scans. 
Finally, the processed scans are combined and reshaped back into image patches by the Scan Merging Module. 
The integration of these modules within the AMS6 Block allows for a more robust and uncertainty-aware approach to image reconstruction, leveraging the strengths of S6 Models while introducing a novel mechanism to account for and estimate uncertainty through arbitrary masking and Monte Carlo dropout techniques.

\subsection{MambaMIR: Model Architecture}

\subsubsection{Overall Architecture.}

The architecture of MambaMIR, depicted in Fig.~\ref{fig:FIG_ARCHITECTURE} (A), embodies a comprehensive framework for medical image reconstruction. The model consists of an Input Module, a series of Arbitrary-Masked State Space (AMSS) Block Groups, and an Output Module, interconnected by residual connections to facilitate effective feature processing and image reconstruction.
The Input Module acts as a stem module, transforming images from the pixel space $\mathbb{R}^{h \times w \times c}$ into a latent patch representation $\mathbb{R}^{H \times W \times C}$. The module effectively increases the dimension from individual pixels to aggregated patch representations, allowing for enhanced feature extraction.
The Output Module inversely maps the processed features from the latent space $\mathbb{R}^{H \times W \times C}$ back to the original pixel space $\mathbb{R}^{h \times w \times c}$, facilitating the reconstruction of high-fidelity medical images.
A cascade of $N$ AMSS Block Groups are placed in the middle for feature processing, each of which contains a convolutional layer, a layer normalisation layer, and $M$ AMSS Blocks.
Residual connection is applied at for both the whole MambaMIR and the cascaded AMSS Blocks, ensuring efficient and stable training.

\subsubsection{AMSS Block.}

The structure of AMSS Block, drawing inspiration from the architecture of the Mamba Block~\cite{Gu2023Mamba} and VSS Block~\cite{Liu2024VMamba}, is shown in Fig.~\ref{fig:FIG_ARCHITECTURE} (B).
Instead of taking the typical structure of ViT Block, i.e., ``Norm-Attn.-Norm-MLP'', AMSS Blocks discard the components of ``-Norm-MLP'' for a lighter network size. 

The input of AMSS Blocks undergoes through a layer normalisation layer, and then is split into two pathways. 
The primary pathway sequentially proceeds through an gating linear layer, a depth-wise convolution layer with a 3x3 kernel followed by a SiLU activation function~\cite{Ramachandran2017Searching}, an AMS6 Block and a layer normalisation layer. 
Concurrently, the secondary pathway is processed through an linear layer with a SiLU activation function.
The outputs of these two pathways are subsequently merged via multiplication and directed through a final gating linear layer to produce the AMSS Block's output. 

\subsection{Optimisation Scheme}
Following the taxonomy in~\cite{Hammernik2022Physics}, the proposed model can be categorised as an end-to-end enhancement-based reconstruction model~\cite{Hammernik2022Physics}, and can be expressed as $\hat x_u = \operatorname{MambaMIR(x_u)}$, where $\hat x_u$ is the reconstructed medical image and $x_u$ is the subsampled image.
To train our proposed MambaMIR, we utilised Charbonnier in both image and transform domains (Fourier domain for MRI and Radon domain for CT), denoted as $\mathcal{L}_{\mathrm{img}}(\theta)$ and $\mathcal{L}_{\mathrm{trans}}(\theta)$ respectively. 
A $l1$ perceptual loss, $\mathcal{L}_{\mathrm{perc}}(\theta)$, is applied in the latent space of a pre-trained VGG $f_{\mathrm{VGG}}(\cdot)$, which can be expressed as:
\begin{equation}\label{eq:loss_list}
\begin{aligned}
&\mathop{\text{min}}\limits_{\theta} 
\mathcal{L}_{\mathrm{img}}(\theta) =
\sqrt{\mid\mid x - \hat x_u \mid\mid^2_2 + \epsilon^2},\\
&\mathop{\text{min}}\limits_{\theta} 
\mathcal{L}_{\mathrm{trans}}(\theta) =
\sqrt{\mid\mid \mathcal{T}x - \mathcal{T} \hat x_u \mid\mid^2_2 + \epsilon^2}, \quad \mathcal{T}:=\mathcal{F} \text{ or } \mathcal{D},\\\
&\mathop{\text{min}}\limits_{\theta} 
\mathcal{L}_{\mathrm{perc}}(\theta) =
\mid\mid f_{\mathrm{VGG}}(x) - f_{\mathrm{VGG}}(\hat x_u) \mid\mid_1,
\end{aligned}
\end{equation} 

\noindent where $x$ is the ground truth. $\epsilon$ is empirically set to $10^{-9}$. We denote $\theta$ as the network parameter of MambaMIR, and $\mathcal{F}$ and $\mathcal{D}$ refers to the discretised Fourier and Radon transform. 
The total loss of MambaMIR, $\mathcal{L}_{\mathrm{MambaMIR}}(\theta)$ can be then expressed as:
\begin{equation}\label{eq:loss_total}
\begin{aligned}
\mathcal{L}_{\mathrm{MambaMIR}}(\theta)
= \alpha \mathcal{L}_{\mathrm{img}}(\theta)
+ \beta \mathcal{L}_{\mathrm{trans}}(\theta)
+ \gamma \mathcal{L}_{\mathrm{perc}}(\theta),
\end{aligned}
\end{equation} 
\noindent where $\alpha$, $\beta$ and $\gamma$ are weighting parameters.

The proposed MambaMIR-GAN takes the MambaMIR as the the generator $G_{\theta_{G}}$ parameterised by $\theta_{G}$, as well as a U-Net based discriminator~\cite{Schonfeld2020UNet}, $D_{\theta_{D}}$, for adversarial training. 
The adversarial loss $\mathcal{L}_{\mathrm{adv}}(\theta_{G}, \theta_{D})$ and the total loss for MambaMIR-GAN are written as:
\begin{equation}\label{eq:loss_gan}
\begin{aligned}
\mathop{\text{min}}\limits_{\theta_{G}} 
\mathop{\text{max}}\limits_{\theta_{D}}
\mathcal{L}_{\mathrm{adv}}(\theta_{G}, \theta_{D}) 
=\mathbb{E}_{x \sim p_{\mathrm{t}}(x)}
[\mathop{\text{log}} D_{\theta_{D}}(x)]
- \mathbb{E}_{x_u \sim p_{\mathrm{u}}(x_u)}
[\mathop{\text{log}} D_{\theta_{D}}(\hat x_u)], \\
\mathcal{L}_{\mathrm{MambaMIR-GAN}}(\theta_{G}, \theta_{D})
= \mathcal{L}_{\mathrm{MambaMIR}}(\theta_{G})
+ \eta \mathcal{L}_{\mathrm{adv}}(\theta_{G}, \theta_{D}),
\end{aligned}
\end{equation} 
\noindent where $\eta$ is the weighting parameter.

\section{Experimental Results}

\subsection{Dataset}


In this work, we utilised 
FastMRI knee dataset~\cite{Zbontar_2018_fastMRI} for fast MRI reconstruction,
alongside two distinct anatomical subsets were utilised from Low-Dose CT Image and Projection Datasets~\cite{moen2021low} for SVCT reconstruction.

\subsubsection{FastMRI.}
For FastMRI dataset~\cite{Zbontar_2018_fastMRI}, 584 scans of three-dimension (3D), 15-coils, proton-density weighted knee MRI without fat suppression with available ground truth were utilised in the experiments section. To avoid including very noisy or void slices, we chose 20 coronal-view 2D complex-value slices near the centre for each cases, and all slices were centre-cropped to $320 \times 320$ in the image space.
All the scans were randomly divided into training set (420 cases, 8400 slices), validation set (64 cases, 1280 slices) and testing set (100 cases, 20 slices), approximately according to a ratio of 7:1:2. 
The emulated single-coil data officially provided was used as the single-coil complex-value ground truth.

\subsubsection{Low-Dose CT Image and Projection Datasets.}
For Low-Dose CT Image and Projection Datasets~\cite{moen2021low}, The chest subset comprised low-dose non-contrast scans aimed at screening high-risk patients for pulmonary nodules, while the abdomen subset consisted of contrast-enhanced CT scans used for detecting metastatic liver lesions. Each subset included 40 cases.
In our experiments, we spilt scans in each subset into training set (32 cases, 12821 slice in chest scans; 32 cases, 5904 slice in abdomen scans) and testing set (8 cases, 3206 slices in chest scans; 8 cases, 1476 slice in abdomen scans). Sparse-view sinograms were generated using torch-radon~\cite{torch_radon} in a fan-beam CT geometry, with 60 projection views and 736 detectors. The source-to-detector distance was set to 1000 mm, and the source-to-rotation-centre distance was 512 mm. The reconstructed image resolution was $512 \times 512$ pixels.

\subsection{Implementation Details and Evaluation Metrics}

\begin{table}[htbp]
  \centering
  \caption{The quantitative results of comparison experiments on FastMRI dataset with acceleration factor (AF) $\times 8$ and $\times 16$.}
  \resizebox{\textwidth}{!}{
    \begin{tabular}{ccccccc}
    \toprule
    \multirow{2}[4]{*}{Method} & \multicolumn{3}{|c}{AF $\times$ 8} & \multicolumn{3}{|c}{AF $\times$ 16} \\
\cmidrule{2-7}          & \multicolumn{1}{|c}{SSIM $\uparrow$}  & PSNR $\uparrow$  & LPIPS $\downarrow$ & \multicolumn{1}{|c}{SSIM $\uparrow$}  & PSNR $\uparrow$  & LPIPS $\downarrow$ \\
    \midrule
    ZF    & 0.482 (0.098) & 22.75 (1.73) & 0.504 (0.058) & 0.415 (0.092) & 20.04 (1.60) & 0.580 (0.049) \\
    D5C5  & 0.548 (0.111) & 25.99 (2.14) & 0.292 (0.039) & 0.474 (0.111) & 23.36 (1.79) & 0.412 (0.049) \\
    DAGAN & 0.530 (0.106) & 25.19 (2.21) & 0.262 (0.043) & 0.487 (0.105) & 23.88 (1.84) & 0.317 (0.045) \\
    SwinMR & 0.568 (0.116) & 26.98 (2.47) & 0.254 (0.043) & 0.496 (0.115) & 24.86 (2.12) & 0.327 (0.046) \\
    STGAN & 0.594 (0.105) & 26.90 (2.31) & \textbf{0.155 (0.040)} & 0.541 (0.102) & 24.86 (1.98) & 0.221 (0.055) \\
    \midrule
    MambaMIR & 0.576 (0.119) & \textbf{27.13 (2.52)} & 0.251 (0.043) & 0.513 (0.117) & \textbf{25.10 (2.10)} & 0.323 (0.042) \\
    MambaMIR-GAN & \textbf{0.600 (0.107)} & 27.09 (2.36) & 0.159 (0.034) & \textbf{0.543 (0.106)} & 25.09 (2.00) & \textbf{0.203 (0.040)} \\
    \bottomrule
    \end{tabular}%
    }
  \label{tab:MRI-FastMRI}%
\end{table}%

\begin{table}[!t]
  \centering
  \caption{The quantitative results of comparison experiments on chest and abdomen subsets of Low-Dose CT Image and Projection datasets with 60 projection views.}
  \resizebox{\textwidth}{!}{
    \begin{tabular}{ccccccc}
    \toprule
    \multirow{2}[3]{*}{Method} & \multicolumn{3}{|c}{Chest} & \multicolumn{3}{|c}{Abdomen} \\
\cmidrule{2-7}          & \multicolumn{1}{|c}{SSIM $\uparrow$}  & PSNR $\uparrow$  & LPIPS $\downarrow$ & \multicolumn{1}{|c}{SSIM $\uparrow$}  & PSNR $\uparrow$  & LPIPS $\downarrow$ \\
    \midrule
    FBP   & 0.550 (0.036) & 28.52 (1.02) & 0.440 (0.037) & 0.716 (0.041) & 31.88 (1.26) & 0.410 (0.035) \\
    Inter & 0.847 (0.025) & 35.56 (1.02) & 0.155 (0.031) & 0.947 (0.008) & 40.96 (0.97) & 0.076 (0.018) \\
    DDnet & 0.835 (0.027) & 35.41 (1.01) & 0.154 (0.023) & 0.941 (0.010) & 40.35 (1.00) & 0.090 (0.016) \\
    FBPconv & 0.801 (0.035) & 34.13 (0.97) & 0.357 (0.050) & 0.929 (0.013) & 38.14 (1.09) & 0.175 (0.025) \\
    IRadonMap & 0.868 (0.029) & 36.84 (1.16) & \textbf{0.135 (0.025)} & 0.968 (0.007) & 43.32 (1.10) & 0.060 (0.015) \\
    Regformer & 0.850 (0.035) & 36.15 (1.14) & 0.226 (0.062) & 0.966 (0.007) & 42.64 (1.13) & 0.077 (0.021) \\
    \midrule
    MambaMIR & \textbf{0.870 (0.043)} & \textbf{37.02 (1.51)} & 0.174 (0.038) & \textbf{0.975 (0.006)} & 43.70 (1.10) & 0.044 (0.015) \\
    MambaMIR-GAN & 0.868 (0.043) & 36.94 (1.50) & 0.151 (0.026) & 0.971 (0.007) & \textbf{44.34 (1.14)} & \textbf{0.035 (0.008)} \\
    \bottomrule
    \end{tabular}%
    }
  \label{tab:SVCT}%
\end{table}%


In the architecture of MambaMIR, we configured $N=6$ AMSS Block Groups, with each group comprising $M=2$ AMSS Blocks. The embedding dimension is set to 180, and the patch size is chosen as 4. 
During training, 2D slice images are randomly cropped to a size of $192 \times 192$ to enhance training efficiency. The loss function incorporates weighting parameters $\alpha$, $\beta$, $\gamma$, and $\eta$ (specifically for MambaMIR-GAN), which are set to 15, 0.1, 0.0025, and 0.1, respectively.

Our proposed models, MambaMIR and MambaMIR-GAN, were trained on one and two NVIDIA RTX 3090 GPUs (24GB each), respectively, and evaluated on a single NVIDIA RTX 3090 GPU. Both models underwent training for 100,000 gradient steps, utilizing the Adam optimiser with a learning rate of $2 \times 10^{-5}$ and a batch size of 8.

For quantitative analysis, we employed Peak Signal-to-Noise Ratio (PSNR), Structural Similarity Index Measure (SSIM), and Learned Perceptual Image Patch Similarity (LPIPS)~\cite{Zhang2018LPIPS} to assess reconstruction quality. Notably, LPIPS, as a deep learning-based perceptual metric, closely aligns with human visual perception, providing a more nuanced evaluation of image reconstruction fidelity.

\subsection{Comparison Studies}

\begin{figure}[ht]
    \centering
    \includegraphics[width=4.5in]{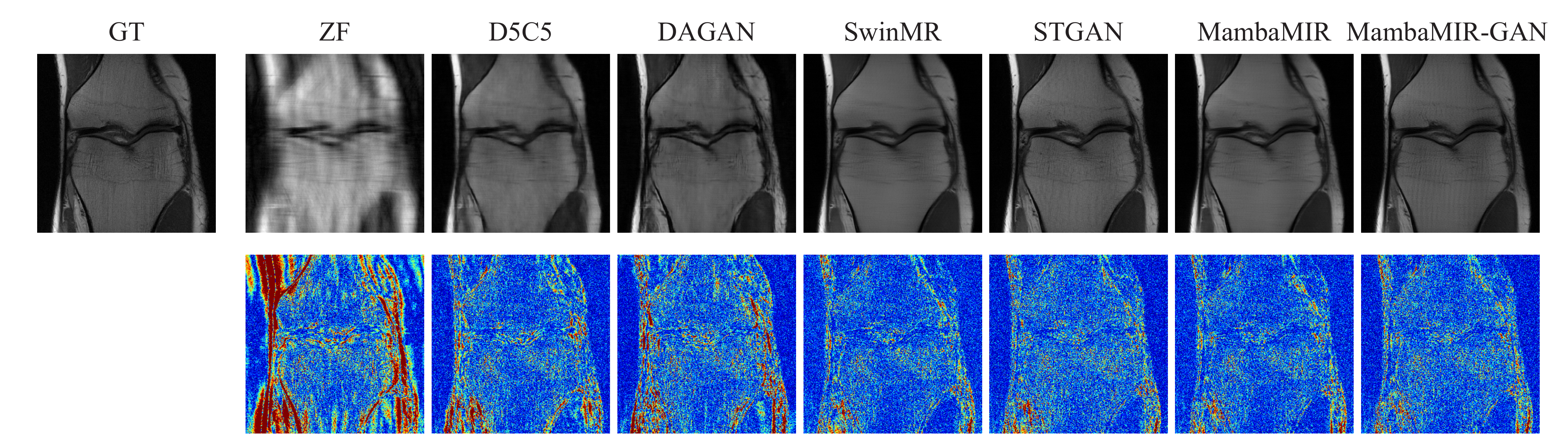}
    \caption{
    The visualised results of comparison experiments on FastMRI dataset with acceleration factor (AF) $\times 8$.
    }
    \label{fig:FIG_VIS_FastMRI_AF8}
\end{figure}

\begin{figure}[ht]
    \centering
    \includegraphics[width=4.5in]{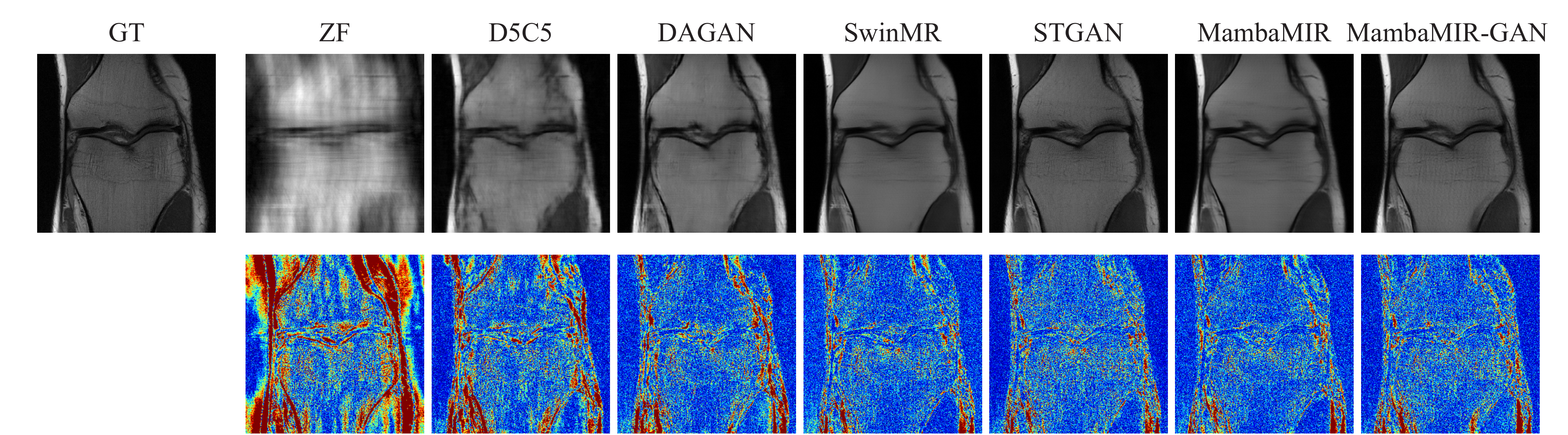}
    \caption{
    The visualised results of comparison experiments on FastMRI dataset with acceleration factor (AF) $\times 16$.
    }
    \label{fig:FIG_VIS_FastMRI_AF16}
\end{figure}

\begin{figure}[ht]
    \centering
    \includegraphics[width=4.5in]{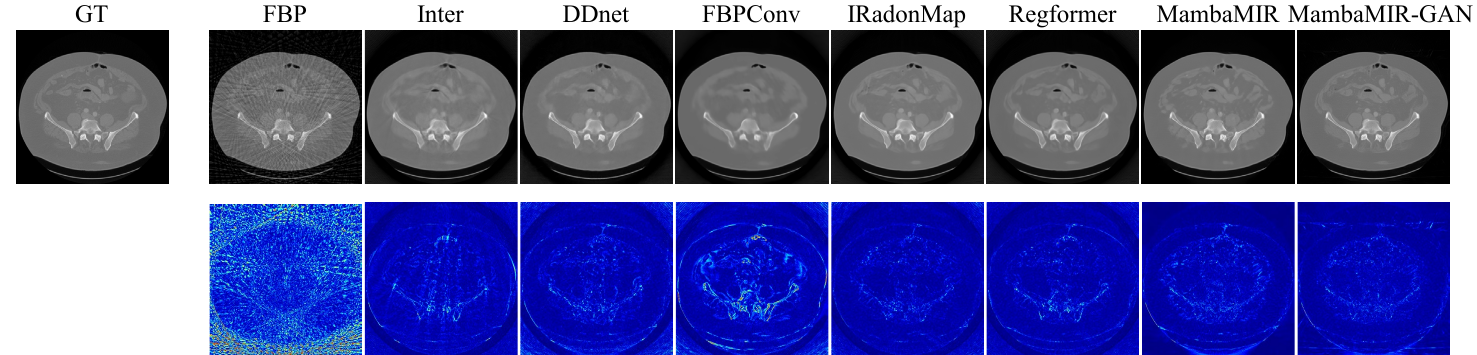}
    \caption{
    The visualised results of comparison experiments on the abdomen subset of Low-Dose CT Image and Projection datasets.
    }
    \label{fig:FIG_VIS_SVCT_Abdomen}
\end{figure}

\begin{figure}[ht]
    \centering
    \includegraphics[width=4.5in]{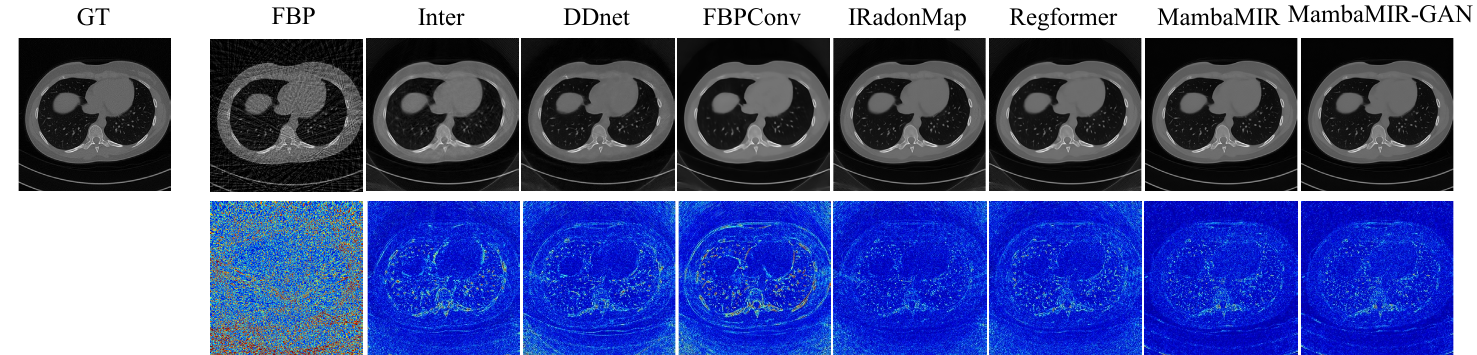}
    \caption{
    The visualised results of comparison experiments on the chest subset of Low-Dose CT Image and Projection datasets.
    }
    \label{fig:FIG_VIS_SVCT_Chest}
\end{figure}

The comparison experimemnt for MRI reconstruction included state-of-the-art methods across D5C5~\cite{Schlemper2017D5C5}, DAGAN~\cite{Yang2018DAGAN}, SwinMR~\cite{Huang2022SwinMR}, STGAN~\cite{Huang2022STGAN}.
The quantitative results is presented in Table~\ref{tab:MRI-FastMRI}, and the visualised samples are shown in Fig.~\ref{fig:FIG_VIS_FastMRI_AF8} and Fig.~\ref{fig:FIG_VIS_FastMRI_AF16}.
For SVCT, we selected state-of-the-art methods for comparison across View Interpolation (Inter)~\cite{lee2017view}, DDNet~\cite{zhang2018sparse}, FBPConv~\cite{jin2017deep}, IRadonMap~\cite{he2020radon} and Regformer~\cite{xia2023transformer}.
The quantitative results is presented in Table~\ref{tab:SVCT}, and the visualised samples are shown in Fig.~\ref{fig:FIG_VIS_SVCT_Abdomen} and Fig.~\ref{fig:FIG_VIS_SVCT_Chest}.
Our proposed MambaMIR and MambaMIR-GAN achieved comparable achieved comparable or superior reconstruction results with respect to state-of-the-art methods. MambaMIR-GAN tends to achieve better perceptual score than MambaMIR, due to the utilisation of the adversarial training strategy.

\begin{figure}[ht]
    \centering
    \includegraphics[width=5in]{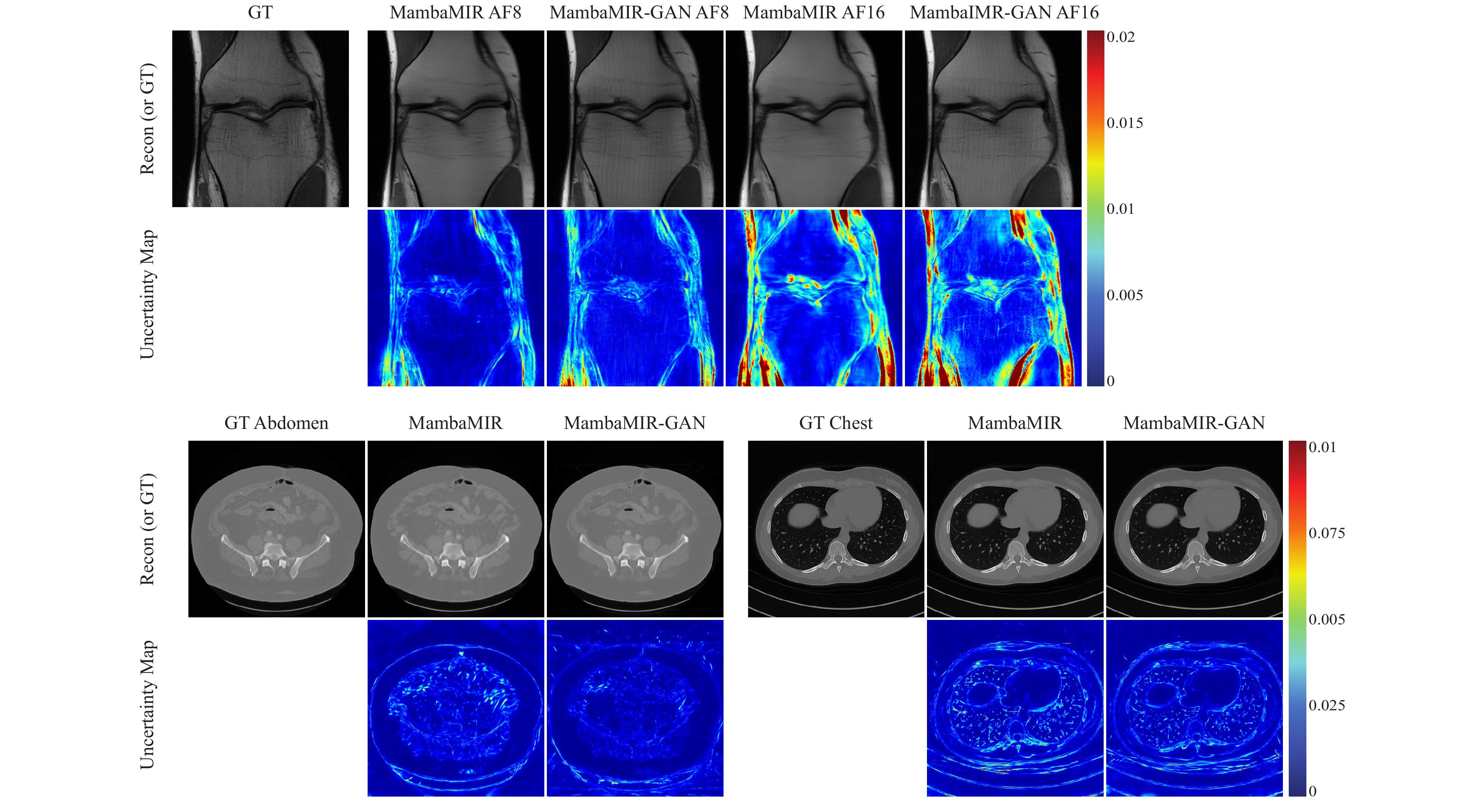}
    \caption{
    The visualised examples of the uncertainty maps on FastMRI dataset and Low-Dose CT Image and Projection datasets.
    }
    \label{fig:FIG_UNC}
\end{figure}

In addition, our MambaMIR further provides an uncertainty metric, offering a visual representation of the confidence in the reconstructed images (Fig.~\ref{fig:FIG_UNC}). 

\section{Discussion and Conclusion}

In this paper, we have proposed the Mamba-based model, MambaMIR, for joint medical image reconstruction, alongside its GAN-based variant, MambaMIR-GAN. Our experiments have demonstrated that MambaMIR achieved reconstruction results that are on par with or superior to current state-of-the-art methods. Furthermore, the estimated uncertainty maps have offered additional insights into the reliability of the reconstructed image quality.

The experimental outcomes have suggested that both MambaMIR and MambaMIR-GAN delivered comparable or superior performance in medical image reconstruction. Notably, MambaMIR-GAN may provide reconstructions that better align with human perceptual qualities.
In MRI reconstruction, MambaMIR and MambaMIR-GAN have attained the highest SSIM and PSNR values for both tested AFs. MambaMIR-GAN has achieved the best performance in terms of LPIPS at AF $\times 16$ and comparably at AF $\times 8$.
Incorporating GAN into MambaMIR-GAN has significantly enhanced the perceptual quality of the images. 
For SVCT reconstruction, MambaMIR and MambaMIR-GAN have surpassed all other methods in both tested subsets, achieving the best SSIM and PSNR scores. MambaMIR-GAN has obtained the best LPIPS on the abdomen subset and comparable results on the chest subset.

Additionally, MambaMIR can generate uncertainty maps through the arbitrary-masked mechanism. These maps have visually represented the model's confidence in the reconstructed images, highlighting areas of potential uncertainty which may signal regions with lower image quality or artefacts.

For MRI reconstruction, the uncertainty maps have shown denser for the higher AF (AF $\times 16$), indicating increased uncertainty in the reconstructed image, which is expected since higher AF imply less \textit{k}-space data and thus more reconstruction challenge.
Comparing the MambaMIR and MambaMIR-GAN methods, the MambaMIR-GAN appears to have slightly higher uncertainty, particularly in areas with rich texture information.
For SVCT reconstruction, the uncertainty maps for both the abdomen and chest CTs show a low level of uncertainty across the images for both MambaMIR and MambaMIR-GAN, with the MambaMIR-GAN showing slightly less uncertainty in certain areas.
This has suggested that for SVCT, the GAN-based approach might be more robust in providing high-confidence reconstructions.

In conclusion, the MambaMIR and MambaMIR-GAN models represent significant advancements in the field of medical image reconstruction. 
Future studies may investigate the scalability of these models to various imaging modalities and their potential in computational efficiency.

%

\bibliographystyle{splncs04}
\bibliography{references.bib}
\end{document}